\documentclass[twocolumn]{jpsj3}
\usepackage{graphicx,color}

\newcommand{\etal}{{\it et al.}\ }

\title{First-Principles Electronic Structure of Superconductor Ca$_4$Al$_2$O$_6$Fe$_2$P$_2$:\\
Comparison with LaFePO and Ca$_4$Al$_2$O$_6$Fe$_2$As$_2$}
\author{Taichi Kosugi$^1$, Takashi Miyake$^{1,2,3}$, and Shoji Ishibashi$^{1,3}$}

\inst{
$^1$Nanosystem Research Institute (NRI) ``RICS'', AIST, 1-1-1 Umezono, Tsukuba 305-8568, Japan \\
$^2$Japan Science and Technology Agency, CREST, 4-1-8 Honcho, Kawaguchi, Saitama 332-0012, Japan \\
$^3$Japan Science and Technology Agency, TRIP, 5 Sanbancho, Chiyoda, Tokyo 102-0075, Japan \\
}

\abst{
We investigate the electronic structures of 
iron-based superconductors having perovskite-like blocking layers, 
Ca$_4$Al$_2$O$_6$Fe$_2$P$_2$ and 
Ca$_4$Al$_2$O$_6$Fe$_2$As$_2$, from first principles. 
Ca$_4$Al$_2$O$_6$Fe$_2$P$_2$ is found to have two hole-like Fermi surfaces around $\Gamma$, 
and one hole-like Fermi surface around M in the unfolded Brillouin zone. 
This is in contrast with LaFePO, where no Fermi surface is found around M. 
The relationship between their band structures and measured transition temperatures of superconductivity is discussed.
The number of Fermi surfaces in Ca$_4$Al$_2$O$_6$Fe$_2$P$_2$ is also 
different from that in Ca$_4$Al$_2$O$_6$Fe$_2$As$_2$, in which 
only one Fermi surface is formed around $\Gamma$. 
Analysis using maximally localized Wannier functions clarifies 
that the differences between their band structures originate mainly from the difference in pnictogen height.
We then analyze the alloying effect on the electronic structure of Ca$_4$Al$_2$O$_6$Fe$_2$AsP.  
It is found that this electronic structure is similar to those of 
Ca$_4$Al$_2$O$_6$Fe$_2$P$_2$ and Ca$_4$Al$_2$O$_6$Fe$_2$As$_2$ with average crystal structures,
though Ca$_4$Al$_2$O$_6$Fe$_2$AsP contains pnictogen height disorder.
We calculate the generalized susceptibility for Ca$_4$Al$_2$O$_6$Fe$_2$(As$_{1-x}$P$_x$)$_2$
and clarify the factors determining its tendency. 
}

\kword{Fe-based superconductor, Ca$_4$Al$_2$O$_6$Fe$_2$P$_2$, Ca$_4$Al$_2$O$_6$Fe$_2$As$_2$, LaFePO, electronic structure, first-principles calculation, maximally localized Wannier function}

\begin{document}
\maketitle

\section{Introduction}

Since the discovery of iron-based superconductors
in LaFePO with its transition temperature $T_\mathrm{c}$ of $5$ K\cite{bib:1988} and in LaFeAsO with $T_\mathrm{c} = 26$ K,\cite{bib:1946}
many attempts to clarify the material properties and to realize higher $T_{\mathrm{c}}$ have been continued,
and Fe-based superconductors with various chemical formulae and crystal structures have been reported,
among which SmFeAsO has shown the highest $T_{\mathrm{c}}$ of $55$ K to date is realized.\cite{bib:SmFeAsO_55K}

An Fe-based superconductor comprises two-dimensional iron (Fe)-anion ($A$) layers. 
The electronic bands near the Fermi level have a strong Fe $d$ character \cite{bib:1183,bib:1956,bib:1918,bib:1783}, 
which should be responsible for the superconductivity. 
Lee \etal\cite{bib:1951} plotted the $T_{\mathrm{c}}$ of various Fe-based superconductors as a function of the $A$-Fe-$A$ bond angle $\alpha$
and demonstrated that the $T_{\mathrm{c}}$ shows a peak at nearly $109.47^\circ$, for which Fe$A_4$ forms a regular tetrahedron.
Much attention has thus been paid to the strong correlation between $T_{\mathrm{c}}$ and the local geometry of an Fe$A_4$ tetrahedron, namely, the lattice constant $a$, the Fe-$A$ bond length $d$, the bond angle $\alpha$, and the anion height $h_A$. 
Two of these four geometry constants are independent.
A small $a$ or $d$ leads to a small density of states at the Fermi level
and is thus expected to be unfavorable for superconductivity. 
On the other hand, a large $h_A$ allows a robust hole Fermi surface around $(\pi, \pi)$,\cite{vildosola} 
which interacts with electron Fermi surfaces
and is thus expected to be favorable for superconductivity  \cite{bib:1957}.
The measured $T_{\mathrm{c}}$'s, however, are not necessarily higher at larger $h_A$'s.
Mizuguchi and coworkers\cite{bib:1958,bib:1748} provided a plot of $T_{\mathrm{c}}$ as a function of $h_A$,
which obeys a symmetric curve with a peak at approximately $1.38$ \AA, common to 1111-, 122-, 111-, and 11-type superconductors.


Since the reports for Sr$_4$Sc$_2$O$_6$Fe$_2$P$_2$\cite{bib:1932} and Sr$_4$V$_2$O$_6$Fe$_2$As$_2$\cite{bib:1943},
Fe-based materials with perovskite-type blocking layers have been known to show superconductivity
for various combinations of their formulae and thickness.\cite{bib:1959,bib:1947,bib:1949,bib:1948,bib:1945,bib:1933,bib:1931,bib:1986,bib:1950}
They include
Ca$_4$Al$_2$O$_6$Fe$_2$As$_2$ ($a = 3.713$ \AA \, and $c = 15.407$ \AA, $\alpha = 102.13^\circ$, and $h_{\mathrm{As}} = 1.500$ \AA) with $T_{\mathrm{c}} = 28.3$ K\cite{bib:1933} and
Ca$_4$Al$_2$O$_6$Fe$_2$P$_2$ ($a = 3.692$ \AA \, and $c = 14.934$ \AA, $\alpha = 109.45^\circ$, and $h_{\mathrm{P}} = 1.306$ \AA) with $T_{\mathrm{c}} = 17.1$ K.\cite{bib:1933}

Among the Fe-based superconductors reported so far,
Ca$_4$Al$_2$O$_6$Fe$_2$As$_2$ and Ca$_4$Al$_2$O$_6$Fe$_2$P$_2$ are of particular interest
since they have rather small $a$'s, owing to the small ion radius of Al$^{3+}$, and Ca$_4$Al$_2$O$_6$Fe$_2$As$_2$ has a rather large $h_{\mathrm{As}}$.
Furthermore, it is important to answer why the $T_{\mathrm{c}}$ of Ca$_4$Al$_2$O$_6$Fe$_2$P$_2$ is higher than that of LaFePO
to elucidate the mechanism of superconductivity in systems consisting of FeP layers.
Our previous work\cite{bib:1605} revealed that $\alpha$
has a strong impact on the band rearrangement and the Fermi surface topology in Ca$_4$Al$_2$O$_6$Fe$_2$As$_2$. 
Subsequently, Usui and Kuroki\cite{bib:1934} pointed out that the superconductivity of an Fe-based superconductor is optimized for $\alpha$ of a regular tetrahedron,
which maximizes the hole Fermi surface multiplicity.

The superconducting properties of Fe-based systems depend sensitively on multiple factors related to the crystal structure and formula, as outlined above.
It is thus important to investigate the electronic properties of such systems with the crystal structure and/or formula varied,
focusing on the behavior of Fermi surfaces and electronic orbitals in the vicinity of the Fermi level.
In the present study, we therefore perform first-principles electronic structure calculations for Ca$_4$Al$_2$O$_6$Fe$_2$(As$_{1-x}$P$_x$)$_2$.
We first compare the band structures of Ca$_4$Al$_2$O$_6$Fe$_2$P$_2$ and Ca$_4$Al$_2$O$_6$Fe$_2$As$_2$,
and determine the origins of the differences in their band structures by comparing the transfer integrals between localized electronic orbitals.
We extract the transfer integrals by constructing maximally localized Wannier functions\cite{bib:MLWF},
which provide a picture of localized electronic orbitals in solid.
The results of this analysis suggest that the band structures of Ca$_4$Al$_2$O$_6$Fe$_2$(As$_{1-x}$P$_x$)$_2$ are determined mainly by crystal structure, not by chemical composition.
We then demonstrate that electronic structure calculation for Ca$_4$Al$_2$O$_6$Fe$_2$AsP corroborates this idea by unfolding its band structure.
We calculate the generalized susceptibility $\chi_0$, which correlates with superconductivity in general, of Ca$_4$Al$_2$O$_6$Fe$_2$(As$_{1-x}$P$_x$)$_2$
and clarify the factors determining the behavior of $\chi_0$.

\section{Computational Details}

We used the computational code QMAS\cite{bib:QMAS} based on the projector augmented-wave method\cite{bib:PAW},
which has been applied to the study of the ground state properties of LaFeAsO\cite{bib:1783,bib:1831}, SrFe$_2$As$_2$\cite{bib:1912}, and Ca$_4$Al$_2$O$_6$Fe$_2$As$_2$\cite{bib:1605}.
The PBE exchange-correlation functional\cite{bib:GGA} within the generalized gradient approximation (GGA) was adopted.
The pseudo wave functions were expanded in plane waves with an energy cutoff of $40$ Ry.
We employed a $12 \times 12 \times 4$ $k$ point mesh for systems with conventional tetragonal cells and an $8 \times 8 \times 4$ $k$ point mesh with supercells.
The electronic band structure in the vicinity of the Fermi level is analyzed
in detail using the maximally localized Wannier function technique\cite{bib:MLWF}.

For the Ca$_4$Al$_2$O$_6$Fe$_2$P$_2$\cite{bib:1933, privcomm} and Ca$_4$Al$_2$O$_6$Fe$_2$As$_2$\cite{bib:1933} crystal structures,
the experimental values of tetragonal lattice constants and atomic coordinates are used in the present work.

In the analyses of the alloying effect on Ca$_4$Al$_2$O$_6$Fe$_2$AsP, we unfold its band dispersion obtained for a supercell,
following the procedures proposed by Ku \etal\cite{bib:1881}.
The unfolded band dispersion is given as a sum of weighted delta functions:
\begin{gather}
	A_{kn, kn} (\omega) = \sum_{J} | \langle kn| KJ \rangle |^2 \delta(\omega - \varepsilon_{KJ}),
	\label{spec}
\end{gather}
where $K$ is the wave vector in the folded Brillouin zone corresponding to the wave vector $k$ in the unfolded Brillouin zone.
$J$ is the band index for the supercell and $n$ denotes the Wannier function in the primitive cell.
$|KJ \rangle$ is the eigenstate of the Kohn-Sham Hamiltonian for the supercell,
whereas $|kn \rangle$ is the Fourier transform of the Wannier function in the primitive cell.
We are able to unfold $20$ Fe $d$ bands to $5$ bands by choosing the proper gauge of the Wannier functions\cite{bib:1918}.

We calculate the generalized susceptibility $\chi_0$ of Ca$_4$Al$_2$O$_6$Fe$_2$P$_2$ and Ca$_4$Al$_2$O$_6$Fe$_2$As$_2$ with gradually varying their crystal structures.
Its expression for a system with time reversal symmetry is given by
\begin{gather}
	\chi_0(\boldsymbol{q}) = \frac{1}{2} \sum_{m,n, \boldsymbol{k}}
		\frac{f_{m \boldsymbol{k}} - f_{n \boldsymbol{k} + \boldsymbol{q}} }{ \varepsilon_{n \boldsymbol{k} + \boldsymbol{q}} - \varepsilon_{m \boldsymbol{k}} },
	\label{chi0}
\end{gather}
where $f_{m \boldsymbol{k}}$ is the occupation number and $\varepsilon_{m \boldsymbol{k}}$ is the energy eigenvalue.
$\chi_0$ reflects the nesting property of the system and is expected to correlate with superconductivity.
We evaluate eq. (\ref{chi0}) as a summation for $40 \times 40 \times 4$ $k$ points
by interpolating the band dispersion using the Wannier functions of Fe $d$ bands.

\section{Results and Discussion}

\subsection{Electronic properties of Ca$_4$Al$_2$O$_6$Fe$_2$P$_2$ in comparison with those of Ca$_4$Al$_2$O$_6$Fe$_2$As$_2$}

\begin{figure}[htbp]
\begin{center}
\includegraphics[keepaspectratio,width=8cm]{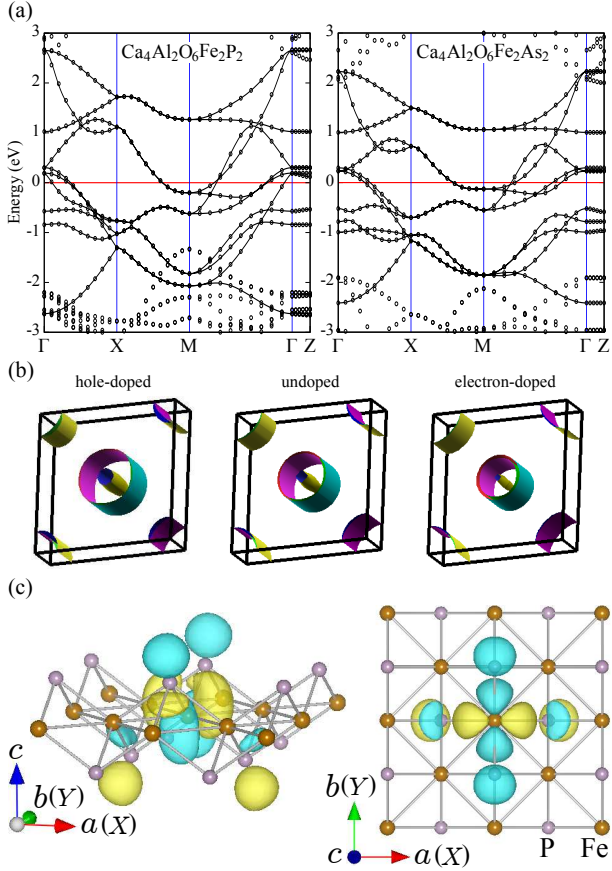}
\end{center}
\caption{
(Color)
(a) Nonmagnetic electronic band structures of Ca$_4$Al$_2$O$_6$Fe$_2$P$_2$ and Ca$_4$Al$_2$O$_6$Fe$_2$As$_2$.
Circles represent DFT-GGA bands and curves represent bands interpolated using the maximally localized Wannier functions of $d$ models.
The origins of energy are set to the respective Fermi levels.
(b) Fermi surfaces for hole-doped (left), undoped (middle), and electron-doped (right) Ca$_4$Al$_2$O$_6$Fe$_2$P$_2$.
(c)
A bird's-eye view of maximally localized Wannier function of Fe $d_{X^2 - Y^2}$ orbital is shown on the left.
The Wannier function viewed along the $c$ axis is shown on the right.
The $a$ and $b$ directions are parallel to the $X$- and $Y$-axes, respectively.
Brown and purple balls represent Fe and P atoms, respectively.
These figures were drawn with VESTA\cite{bib:VESTA}.
}
\label{DFTbands}
\end{figure}

The nonmagnetic electronic band structures of Ca$_4$Al$_2$O$_6$Fe$_2$P$_2$ and Ca$_4$Al$_2$O$_6$Fe$_2$As$_2$
obtained with the first-principles calculation are shown as circles in Fig. \ref{DFTbands}(a).
In each of the systems, the unit cell contains two Fe atoms
and ten bands having a strong Fe $d$ character lie near the Fermi level.
The most prominent difference in the calculated band structure between these two systems is the multiplicity of Fermi surfaces around $\Gamma$:
Ca$_4$Al$_2$O$_6$Fe$_2$P$_2$ has three, whereas Ca$_4$Al$_2$O$_6$Fe$_2$As$_2$ has two. 
We note here that
LaFePO has two Fermi surfaces around $\Gamma$,
which would explain why Ca$_4$Al$_2$O$_6$Fe$_2$P$_2$  has a higher $T_c$ than LaFePO. 
Figure \ref{DFTbands} (b) shows the Fermi surfaces of Ca$_4$Al$_2$O$_6$Fe$_2$P$_2$ in an undoped case and doped cases with $\pm 0.1$ electrons per Fe atom
within the rigid band approximation.
Another noticeable difference is seen in the height of pnictogen ($Pn$) $p$ bands at M.
The top of P $p$ bands in Ca$_4$Al$_2$O$_6$Fe$_2$P$_2$ lies close to $-1.3$ eV at M,
whereas that of As $p$ bands is not at M and those bands are below $-1.9$ eV.

\begin{figure}[htbp]
\begin{center}
\includegraphics[keepaspectratio,width=8cm]{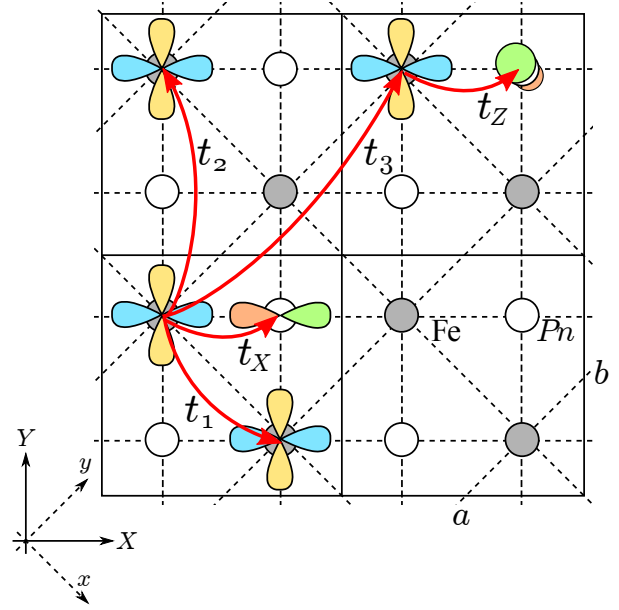}
\end{center}
\caption{
(Color online)
Schematic illustration of transfer integrals of a $dpp$ model.
Solid lines represent unit cells.
Fe $d_{X^2 - Y^2}$ orbitals and
$Pn$ $p_X$ and $p_Z$ orbitals
are shown.
We define the $x$- and $y$-axes by rotating the original $X$- and $Y$-axes by $45^\circ$.
Transfer integrals are shown as red arrows.
$t_i (i=1,2,3)$ is that between the $d_{X^2-Y^2}$ orbitals at the $i$-th nearest neighbors,
while $t_j (j=X,Z)$ is that between the $p_j$ orbital and its nearest-neighboring $d_{X^2-Y^2}$ orbital.
}
\label{transfers}
\end{figure}

To analyze the differences between the two compounds, 
the Wannier function technique is useful, since it enables us to 
evaluate the effective parameters of low-energy electronic states\cite{miyake_crpa}. 
We constructed maximally localized Wannier functions from the ten Fe $d$ bands ($d$ model)
and plotted the interpolated band dispersion as curves in Fig. \ref{DFTbands}(a).
It is found that the Wannier functions accurately reproduce the DFT-GGA band structures near the Fermi level.
The Wannier function of the Fe $d_{X^2 - Y^2} (d_{xy})$ orbital in Ca$_4$Al$_2$O$_6$Fe$_2$P$_2$ is shown in Fig. \ref{DFTbands}(c).
[The $X$- and $Y$-axes refer to the conventional unit cell containing two Fe atoms, whereas the rotation of them by $45^\circ$ defines the $x$- and $y$-axes. (see Fig. \ref{transfers})]
Although the shape of this Wannier function is mainly of the $d_{X^2 - Y^2}$ character,
it involves contributions from P $p$ orbitals.
We observed that the positive (blue) lobe of the $d_{X^2 - Y^2}$ orbital is pushed down the $c$ direction
to hybridize with the P $p$ orbitals below the Fe layer,
while the negative (yellow) lobe is pushed up to hybridize with the P $p$ orbitals above,
as shown in Fig. \ref{DFTbands}(c).
We confirmed that the shape of the corresponding Wannier function in Ca$_4$Al$_2$O$_6$Fe$_2$As$_2$ (not shown) has smaller contributions from the As $p$ orbitals.
This reflects 
(i) the Fe-P distance, which is smaller than the Fe-As distance, and 
(ii) the energy separation of the Fe $d$ and the As $p$ bands in Ca$_4$Al$_2$O$_6$Fe$_2$As$_2$, which is larger than that of the Fe $d$ and the P $p$ bands in Ca$_4$Al$_2$O$_6$Fe$_2$P$_2$.

\begin{figure}[htbp]
\begin{center}
\includegraphics[keepaspectratio,width=8cm]{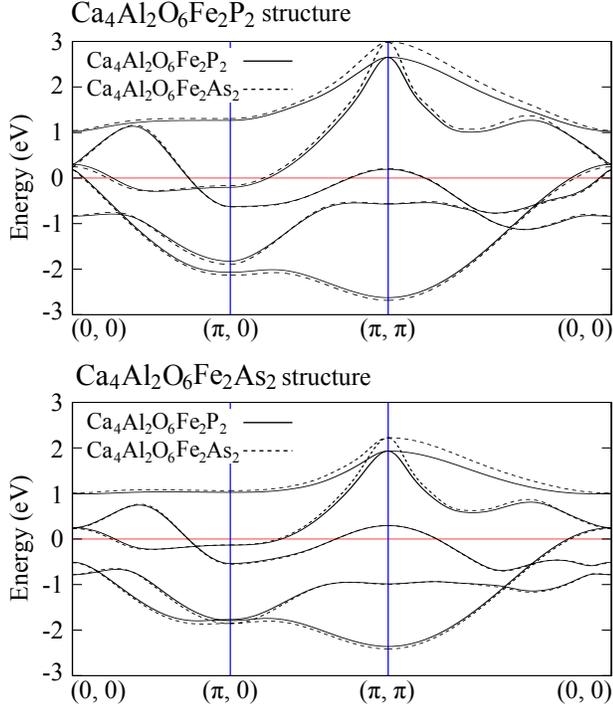}
\end{center}
\caption{
(Color online)
Unfolded $5$-band structures of $d$ models for Ca$_4$Al$_2$O$_6$Fe$_2$P$_2$ and Ca$_4$Al$_2$O$_6$Fe$_2$As$_2$ as solid and dashed curves, respectively.
The upper (lower) panel shows the band structures obtained using the crystal structure of Ca$_4$Al$_2$O$_6$Fe$_2$P$_2$ (Ca$_4$Al$_2$O$_6$Fe$_2$As$_2$).
The origins of energy are set to the respective Fermi levels.
}
\label{unfolded_1}
\end{figure}

To simplify the analysis of the electronic band structure,
we unfolded the interpolated $10$-band dispersion of $d$ models following the procedure proposed by Kuroki \etal\cite{bib:1918}.
The resultant $5$-band structures obtained with the lattice parameters and the atomic coordinates fixed, while varying the chemical composition of the system, are shown in Fig. \ref{unfolded_1}.
It is found that the overall band structures with the same crystal structure look quite similar, especially in the vicinity of the Fermi level.
For the Ca$_4$Al$_2$O$_6$Fe$_2$P$_2$ (Ca$_4$Al$_2$O$_6$Fe$_2$As$_2$) crystal structure, there exist three (two) Fermi surfaces around $\Gamma$ in the original Brillouin zone irrespective of chemical composition,
which correspond to the two (one) $\alpha$ hole pockets  around $(0, 0)$ and one $\gamma$ hole pocket around $(\pi, \pi)$, as shown in Fig. \ref{unfolded_1}.
The numbers of Fermi surfaces around $(0, 0)$ and $(\pi, \pi)$ in the unfolded Brillouin zone for Ca$_4$Al$_2$O$_6$Fe$_2$As$_2$ and Ca$_4$Al$_2$O$_6$Fe$_2$P$_2$ with their own crystal structures are summarized in Table \ref{table_FS}.
We observe in Fig. \ref{unfolded_1} that
the calculated band widths of $d$ models for Ca$_4$Al$_2$O$_6$Fe$_2$As$_2$ with the Ca$_4$Al$_2$O$_6$Fe$_2$As$_2$ and Ca$_4$Al$_2$O$_6$Fe$_2$P$_2$ crystal structures are $4.64$ and $5.68$ eV, respectively.
Those for Ca$_4$Al$_2$O$_6$Fe$_2$P$_2$  with the Ca$_4$Al$_2$O$_6$Fe$_2$As$_2$ and Ca$_4$Al$_2$O$_6$Fe$_2$P$_2$ crystal structures are $4.29$ and $5.28$ eV, respectively.
The Ca$_4$Al$_2$O$_6$Fe$_2$P$_2$ crystal structure gave a larger band width than the Ca$_4$Al$_2$O$_6$Fe$_2$As$_2$ crystal structure when the same chemical composition was used,
since the former has a smaller $a$ and a smaller $h_{Pn}$.
On the other hand, the Ca$_4$Al$_2$O$_6$Fe$_2$P$_2$ composition gave a smaller band width than the Ca$_4$Al$_2$O$_6$Fe$_2$As$_2$ composition when the same crystal structure was used,
since the ion radius of the P atom is smaller than that of the As atom.

\begin{table}[h]
\begin{center}
\caption{
Numbers of Fermi surfaces around $(0, 0)$ and $(\pi, \pi)$ in unfolded Brillouin zone for
Ca$_4$Al$_2$O$_6$Fe$_2$As$_2$ and Ca$_4$Al$_2$O$_6$Fe$_2$P$_2$.
Those for LaFeAsO and LaFePO are also shown for comparison.
}
\label{table_FS}
\begin{tabular}{ccc}
\hline\hline
 & $(0, 0)$ & $(\pi, \pi)$ \\
\hline
Ca$_4$Al$_2$O$_6$Fe$_2$As$_2$ & $1$ & $1$ \\
Ca$_4$Al$_2$O$_6$Fe$_2$P$_2$ & $2$ & $1$ \\
LaFeAsO & $2$ & $1$ \\
LaFePO & $2$ & $0$ \\
\hline\hline
\end{tabular}
\end{center}
\end{table}

It was demonstrated in our previous work\cite{bib:1605} that the Fe $d_{X^2 - Y^2}$ ($d_{xy}$) orbital plays an important role in the band rearrangement of Ca$_4$Al$_2$O$_6$Fe$_2$As$_2$ when the local geometry of the Fe$Pn_4$ tetrahedron is varied.
For the quantitative analysis of the electronic structure of Ca$_4$Al$_2$O$_6$Fe$_2$P$_2$ along this line,
we constructed $dpp$ models of the four systems mentioned above and examined how their chemical composition and crystal structures affect the electronic band structures.
Each of the $dpp$ models consists of $10$ Fe $d$, $6$ $Pn$ $p$, and $18$ O $p$ orbitals.
The transfer integrals and orbital energies of the $d_{X^2 - Y^2}$ orbitals, which are equivalent to the $xy$ orbitals,
and the $Pn$ $p$ orbitals are shown in Table \ref{table_transfer}.
Although $|t_1| \gg |t_2|$ in the $dpp$ models, 
$t_1$ in the $d$ models are small in magnitude;
consequently,
$t_1$ and $t_2$ are comparable in the $d$ models. 
Although the Wannier functions in the $d$ model are spatially extended and have considerable weight at the pnictogen site, 
the $d$ Wannier functions in the $dpp$ model are localized and atomic-orbital-like\cite{aichhorn,miyake}. 
Hence, we can say 
that $t_1$ in the $dpp$ model contains only direct hopping between the neighboring Fe sites,  
whereas $t_1$ in the $d$ model contains indirect hopping via the pnictogen site as well \cite{bib:1605}. 
This indicates that the indirect electron hopping of Fe-$Pn$-Fe partly cancels the direct hopping between the nearest-neighboring $d_{X^2 - Y^2}$ orbitals. 
In the comparison of $dpp$ $t$'s for the same chemical composition but different crystal structures,
the difference between $t_i (i=1, 2)$'s is small compared with their amplitudes.
On the other hand, $t_X$ and $t_Z$ are much larger in the Ca$_4$Al$_2$O$_6$Fe$_2$P$_2$ crystal structure than in the Ca$_4$Al$_2$O$_6$Fe$_2$As$_2$ crystal structure for both chemical compositions.
This is because the hybridization between the $d_{X^2 - Y^2}$ and $p$ orbitals is stronger for a smaller $h_{Pn}$.
Since $t_X$ and $t_Z$ in the $dpp$ models can be regarded to be incorporated into $t$'s in the $d$ models,
the differences in band dispersion between the systems with the same chemical composition but different crystal structures come from the differences between the indirect hopping of Fe-$Pn$-Fe.
The two-dimensional shape of the $d_{X^2 - Y^2}$ orbital allows a larger hybridization with the $p_Z$ orbital than with the $p_X$ orbital,
leading to $t_Z$ much larger than $t_X$ for all the four systems.
In the comparison of $dpp$ $t$'s for the same crystal structure but different chemical compositions,
those for Ca$_4$Al$_2$O$_6$Fe$_2$As$_2$ are basically larger than those for Ca$_4$Al$_2$O$_6$Fe$_2$P$_2$ owing to the ion radius of the As atom larger than that of the P atom.
$t_X$ is, however, larger in Ca$_4$Al$_2$O$_6$Fe$_2$P$_2$ than in Ca$_4$Al$_2$O$_6$Fe$_2$As$_2$ if their crystal structures are the same.
This is probably because the $p$ electrons of a $Pn$ atom are depleted for the hybridization between the $p_Z$ orbital and its neighboring $d_{X^2 - Y^2}$ orbitals.

As clarified above, the differences between the band structures of Ca$_4$Al$_2$O$_6$Fe$_2$P$_2$ and Ca$_4$Al$_2$O$_6$Fe$_2$As$_2$ with their own crystal structures, shown in Fig. \ref{DFTbands}(a),
originate mainly not from the difference in chemical composition, but from that in pnictogen height.
By defining the contribution $t_1^{\mathrm{ind}}$ from the indirect hopping Fe-$Pn$-Fe as the difference between the $t_1$'s of the $d$ and $dpp$ models ($t_1^{\mathrm{ind}} \equiv t_1^d - t_1^{dpp}$),
we can estimate $t_1^{\mathrm{ind}}$ to be $0.462$ eV in Ca$_4$Al$_2$O$_6$Fe$_2$P$_2$ and $0.337$ eV in Ca$_4$Al$_2$O$_6$Fe$_2$As$_2$.
In our previous work\cite{bib:1605}, we estimated $t_1^{\mathrm{ind}}$ for LaFeAsO ($h_{\mathrm{As}} = 1.319$ \AA) to be $0.394$ eV.
These values are reasonable because a smaller $h_{Pn}$ gives a larger $t_1^{\mathrm{ind}}$ for these three systems.
The direct hopping of LaFeAsO, $t_1^{dpp} = -0.243$ eV, is much smaller than those of Ca$_4$Al$_2$O$_6$Fe$_2$P$_2$ and Ca$_4$Al$_2$O$_6$Fe$_2$As$_2$,
which is due to the larger $a$ of LaFeAsO.

\begin{table}[h]
\begin{center}
\caption{Orbital energies and transfer integrals (in eV) of Fe $d_{X^2-Y^2}$ and $Pn$ $p$ Wannier orbitals of $dpp$ models.
The descriptions P and As indicate Ca$_4$Al$_2$O$_6$Fe$_2$P$_2$ and Ca$_4$Al$_2$O$_6$Fe$_2$As$_2$, respectively, for the chemical composition and crystal structure.
Transfer integrals obtained from $d$ models are also shown in parentheses.
The origins of orbital energies $\varepsilon$ are set to the respective Fermi levels.
}
\label{table_transfer}
\begin{tabular}{cccc}
\hline\hline
 &  & \multicolumn{2}{c}{Crystal structure} \\
Composition &  & P & As \\
\hline
 & $\varepsilon_{X^2-Y^2}$ & $-0.977 (0.023)$ & $-0.854 (-0.204)$ \\
 & $t_1$ & $-0.370 (0.092)$ & $-0.362 (-0.044)$ \\
 & $t_2$ & $-0.080 (0.119)$ & $-0.080 (0.065)$ \\
\raisebox{-1.8ex}[0pt][0pt]{P} & $t_3$ & $-0.014 (-0.025)$ & $-0.011 (-0.009)$ \\
 & $\varepsilon_X$ & $-2.491$ & $-2.383$ \\
 & $\varepsilon_Z$ & $-2.537$ & $-2.846$ \\
 & $t_X$ & $0.153$ & $0.069$ \\
 & $t_Z$ & $-0.646$ & $-0.537$ \\
\hline
 & $\varepsilon_{X^2-Y^2}$ & $-1.000 (0.047)$ & $-0.882 (-0.197)$ \\
 & $t_1$ & $-0.387 (0.102)$ & $-0.374 (-0.037)$ \\
 & $t_2$ & $-0.089 (0.121)$ & $-0.087 (0.067)$ \\
\raisebox{-1.8ex}[0pt][0pt]{As} & $t_3$ & $-0.015 (-0.030)$ & $-0.011 (-0.012)$ \\
 & $\varepsilon_X$ & $-2.496$ & $-2.391$ \\
 & $\varepsilon_Z$ & $-2.566$ & $-2.889$ \\
 & $t_X$ & $0.128$ & $0.053$ \\
 & $t_Z$ & $-0.666$ & $-0.552$ \\
\hline\hline
\end{tabular}
\end{center}
\end{table}

\subsection{Alloying effect on Ca$_4$Al$_2$O$_6$Fe$_2$AsP}

\begin{figure}[htbp]
\begin{center}
\includegraphics[keepaspectratio,width=8cm]{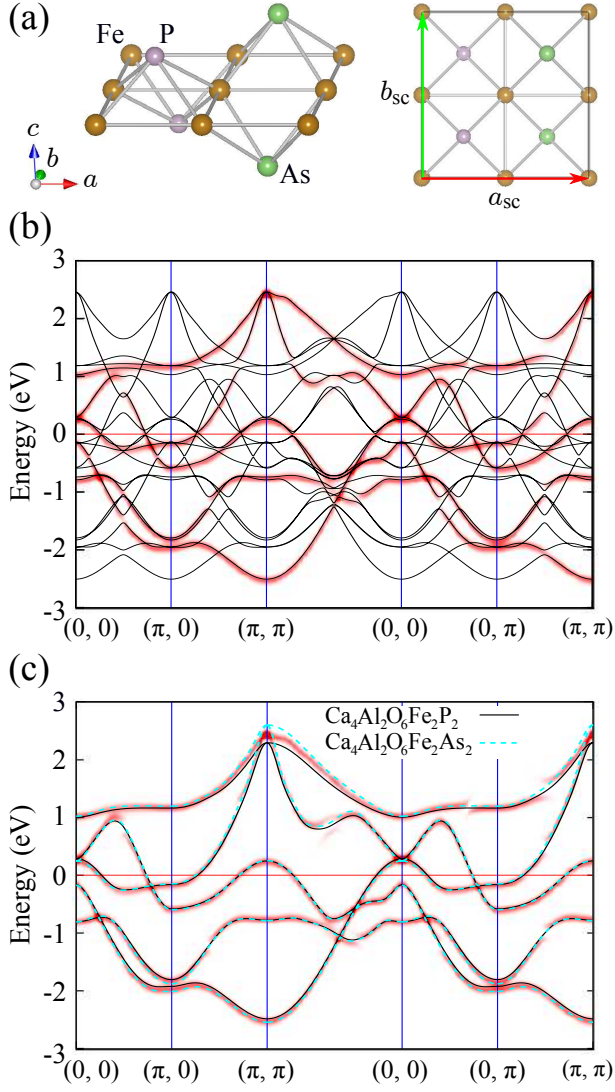}
\end{center}
\caption{
(Color)
(a) Bird's-eye view of an Fe layer of a $\sqrt{2} \times \sqrt{2} \times 1$ supercell for an $x=0.5$ structure is shown on the left.
The layer viewed along the $c$-axis is shown on the right, together with the primitive lattice vectors of the supercell.
Brown, purple, and green balls represent Fe, P, and As atoms, respectively.
(b) The blurred curves are for the unfolded $20$ bands of the $d$ model for the $x=0.5$ system using the supercell.
The solid curves represent the unfolded bands with weights of delta functions fixed constant.
The coordinates of high-symmetry points are with respect to the primitive reciprocal lattice vectors of the primitive cell.
(c) Unfolded band structures of $x=0, 1$, and $0.5$ systems.
The solid and dashed curves are for the $5$-band structures of the $x=1$ and $0$ systems, respectively,
with the crystal structure fixed at the average of Ca$_4$Al$_2$O$_6$Fe$_2$P$_2$ and Ca$_4$Al$_2$O$_6$Fe$_2$As$_2$.
The origins of energy are set to the respective Fermi levels.
}
\label{unfolded_2}
\end{figure}

To study whether or not even the electronic structure of a system in which P and As coexist is determined mainly by crystal structure,
we unfolded the band structures and compared them in the three cases of Ca$_4$Al$_2$O$_6$Fe$_2$(As$_{1-x}$P$_x$)$_2$:
Ca$_4$Al$_2$O$_6$Fe$_2$P$_2$ ($x=1$), Ca$_4$Al$_2$O$_6$Fe$_2$As$_2$ ($x=0$), and Ca$_4$Al$_2$O$_6$Fe$_2$AsP ($x=0.5$).
For the $x=1$ and $0$ systems, the average crystal structure of Ca$_4$Al$_2$O$_6$Fe$_2$P$_2$ and Ca$_4$Al$_2$O$_6$Fe$_2$As$_2$ was used.
The fractional height of the pnictogen atoms used is thus $z_{Pn} = 0.0924$.
For the $x=0.5$ system, we used a $\sqrt{2} \times \sqrt{2} \times 1$ supercell containing four Fe atoms [see Fig. \ref{unfolded_2}(a)].
The lattice parameters and atomic coordinates except P and As were set to the same as those for the $x=1$ and $0$ systems.
The P and As atoms were set at the experimental heights in Ca$_4$Al$_2$O$_6$Fe$_2$P$_2$ ($h_{\mathrm{P}} = 1.306$ \AA) and Ca$_4$Al$_2$O$_6$Fe$_2$As$_2$ ($h_{\mathrm{As}} = 1.500$ \AA), respectively.
The fractional height of the pnictogen atoms used are thus $z_{\mathrm{P}} = 0.0861$ and $z_{\mathrm{As}} = 0.0989$.
The dimensions $\sqrt{2} \times \sqrt{2} \times 1$ of the supercell are the smallest needed for the realization of equal distributions of the two atomic species above and below the Fe layer.
We unfolded the Fe $d$ bands of the $x=0.5$ system using eq. (\ref{spec}).
Specifically, we first constructed $20$ maximally localized Wannier functions of the Fe $d$ character
and then unfolded the interpolated band structure by regarding the supercell as consisting of four primitive cells,
each of which contains one Fe site: $\boldsymbol{a}_\mathrm{sc} = 2 \boldsymbol{a}_\mathrm{pc}, \boldsymbol{b}_\mathrm{sc} = 2 \boldsymbol{b}_\mathrm{pc}$, and $\boldsymbol{c}_\mathrm{sc} = \boldsymbol{c}_\mathrm{pc}$.
We provide in Fig. \ref{unfolded_2}(b) the unfolded band structure for the $x=0.5$ system as blurred curves
and that obtained by fixing the weights of the delta functions in eq. (\ref{spec}) constant as solid curves.
It is clearly seen that the unfolded band structure is essentially 20-fold and 
that the weights of the delta functions, however, make five of them noticeable for almost all the wave vectors.
The unfolded band dispersion of the $x=0.5$ system exhibits different features along the two paths,
one from $(\pi, 0)$ to $(\pi, \pi)$ (corresponding to wave vectors along the $b$ direction) and the other from $(0, \pi)$ to $(\pi, \pi)$ (corresponding to wave vectors along the $a$ direction),
owing to the inequivalence of the $a$ and $b$ directions [see Fig.\ref{unfolded_2}(a)].
Noticeable gap openings of three of the unfolded bands are seen on the latter path,
while the continuity of the unfolded bands on the former path looks firmer.
They originate from the difference between
the arrangement of atoms of different species along the $a$ direction,
and that of the same species along the $b$ direction.

Figure \ref{unfolded_2}(c) shows the unfolded band structures for the $x=0,1$, and $0.5$ systems.
Not only the unfolded band structures for the $x=0$ and $1$ systems, for which the exact unfolding is possible,
but also that for the $x=0.5$ system all look quite similar, especially in the vicinity of the Fermi levels, as well as in Fig. \ref{unfolded_1}.
We also confirmed that the band structure of
Ca$_4$Al$_2$O$_6$Fe$_2$P$_2$ (Ca$_4$Al$_2$O$_6$Fe$_2$As$_2$) with an average crystal structure except for the P (As) atom with its experimental height
has three (two) Fermi surfaces around $\Gamma$.
This indicates that pnictogen height is a crucial factor for the multiplicity of Fermi surfaces around $\Gamma$.
It is interesting to see the proximity of the Fermi level at $(0, 0)$ for the $x=0.5$ system [see Fig. \ref{unfolded_2}(c)].
The top of the second-lowest unfolded band
sits $0.2$ eV below the Fermi level, which is well reproduced by the $x=0$ and $1$ systems with an average crystal structure
despite the $x=0.5$ system containing pnictogen height disorder.
There is an atomic configuration for the chemical composition $x = 0.5$ using the conventional unit cell (see Fig. \ref{transfers}),
in which the atoms of one of the two species are all located above the Fe layer and those of the other species below the layer.
We also calculated the electronic band structure of this system (not shown) with the experimental pnictogen heights of the P and As atoms,
and found that its unfolded bands near the Fermi level are accurately reproduced by the $x=0$ and $1$ systems, as shown in Fig. \ref{unfolded_2}(c).
These results corroborate the independence 
of the electronic structure of a Ca$_4$Al$_2$O$_6$Fe$_2$(As$_{1-x}$P$_x$)$_2$ system
of the chemical composition.
We expect an accurate calculation of a band structure near the Fermi level for an arbitrary $x$ using a crystal structure modeled simply by linearly connecting the Ca$_4$Al$_2$O$_6$Fe$_2$P$_2$ and Ca$_4$Al$_2$O$_6$Fe$_2$As$_2$ crystal structures.

\subsection{Generalized susceptibility of Ca$_4$Al$_2$O$_6$Fe$_2$(As$_{1-x}$P$_x$)$_2$}

\begin{figure}[htbp]
\begin{center}
\includegraphics[keepaspectratio,width=8cm]{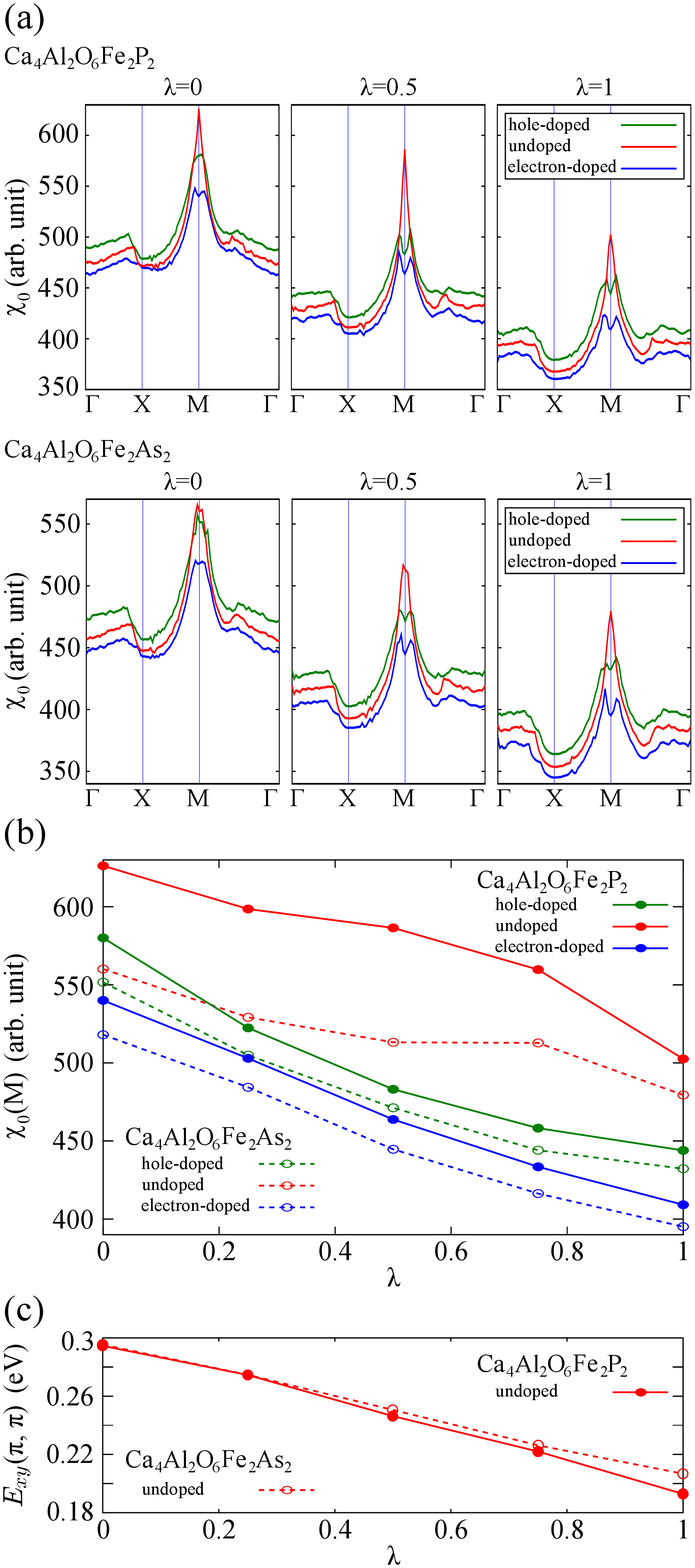}
\end{center}
\caption{
(Color)
(a) Generalized susceptibilities $\chi_0 (\boldsymbol{q})$ of Ca$_4$Al$_2$O$_6$Fe$_2$P$_2$ and Ca$_4$Al$_2$O$_6$Fe$_2$As$_2$ for crystal structure ratios $\lambda = 0, 0.5$ and $1$.
Those in doped cases are also shown.
(b) $\chi_0$(M) in doped and undoped cases of Ca$_4$Al$_2$O$_6$Fe$_2$P$_2$ and Ca$_4$Al$_2$O$_6$Fe$_2$As$_2$ as functions of $\lambda$.
(c) Band energies of $d_{X^2 - Y^2}$-derived bands at $(\pi, \pi)$ for undoped systems.
}
\label{chi0_M}
\end{figure}

Given the results obtained above,
we calculated the generalized susceptibility $\chi_0$ of Ca$_4$Al$_2$O$_6$Fe$_2$P$_2$ and Ca$_4$Al$_2$O$_6$Fe$_2$As$_2$ with gradually varying their crystal structures.
We expect that the  $\chi_0$ calculated in such a way accurately  reproduces the qualitative behavior of Ca$_4$Al$_2$O$_6$Fe$_2$(As$_{1-x}$P$_x$)$_2$ for arbitrary $x$.
The lattice parameters and atomic coordinates $c$ used are  linearly parametrized by the crystal structure ratio $\lambda$ as
$c(\lambda) = \lambda c_{\mathrm{P}} +  (1 - \lambda) c_{\mathrm{As}}$.
$\lambda = 0$ and $1$ correspond to the Ca$_4$Al$_2$O$_6$Fe$_2$As$_2$ and Ca$_4$Al$_2$O$_6$Fe$_2$P$_2$ crystal structures, respectively.
We calculated $\chi_0(\boldsymbol{q})$ using eq. (\ref{chi0}) and 
plotted it in Fig. \ref{chi0_M}(a) for Ca$_4$Al$_2$O$_6$Fe$_2$As$_2$ and  Ca$_4$Al$_2$O$_6$Fe$_2$P$_2$ with $\lambda = 0, 0.5$, and $1$.
Those in doped cases with $\pm 0.1$ electrons per Fe atom with the rigid band approximation are also shown.
The plotted $\chi_0$ is seen to be sensitive to the difference in chemical composition compared with the band energy.
The peaks of $\chi_0$ are located at M in the undoped cases of both Ca$_4$Al$_2$O$_6$Fe$_2$As$_2$ and  Ca$_4$Al$_2$O$_6$Fe$_2$P$_2$,
while, in the doped cases, the peaks are located at incommensurate $\boldsymbol{q}$'s close to M.
We observe that the heights of the peaks are reduced by doping for both Ca$_4$Al$_2$O$_6$Fe$_2$As$_2$ and  Ca$_4$Al$_2$O$_6$Fe$_2$P$_2$,
which suggests that their magnetic instability is suppressed under doping.
Figure \ref{chi0_M}(b) shows the $\chi_0$(M)'s of Ca$_4$Al$_2$O$_6$Fe$_2$P$_2$ and Ca$_4$Al$_2$O$_6$Fe$_2$As$_2$ as functions of $\lambda$.
Although the Ca$_4$Al$_2$O$_6$Fe$_2$P$_2$ systems with smaller band widths give a larger $\chi_0$(M)  than the Ca$_4$Al$_2$O$_6$Fe$_2$As$_2$ systems,
the qualitative behavior of monotonic decrease in $\chi_0$(M) for both Ca$_4$Al$_2$O$_6$Fe$_2$P$_2$ and Ca$_4$Al$_2$O$_6$Fe$_2$As$_2$ are found to be the same in the entire range of $\lambda$.
We found that, with increasing $\lambda$, the number of the $\alpha$ Fermi surfaces around $(0, 0)$ increases from one to two near $\lambda = 0.75$ for both Ca$_4$Al$_2$O$_6$Fe$_2$P$_2$ and Ca$_4$Al$_2$O$_6$Fe$_2$As$_2$ .
The larger number of Fermi surfaces for $\lambda > 0.75$ would be preferable to larger $\chi_0$ because of the larger number of scattering channels.
The calculated $\chi_0$, however, exhibits the trend of a monotonic decrease.
To examine the factors determining this trend, we plot the band energies $E_{xy} (\pi, \pi)$ in Fig. \ref{chi0_M}(c) for the undoped systems as functions of $\lambda$.
$E_{xy} (\pi, \pi)$ is the height of the $d_{X^2 - Y^2}$-derived band.
As is shown, a smaller $\lambda$ gives a higher $E_{xy} (\pi, \pi)$ and thus a more effective $\gamma$ Fermi surface,
and the difference in chemical composition has only a minor effect on the values of $E_{xy} (\pi, \pi)$.
Furthermore, a smaller $\lambda$ (a larger $a$ and a larger $h_{Pn}$) leads to narrower bands and a larger density of states.
These observations suggest that the effects of a small $\lambda$ overcome the difference in the number of $\alpha$ Fermi surfaces.

%

\section{Conclusions}

We performed first-principles calculations of the electronic structures of Ca$_4$Al$_2$O$_6$Fe$_2$P$_2$ and Ca$_4$Al$_2$O$_6$Fe$_2$As$_2$.
Ca$_4$Al$_2$O$_6$Fe$_2$P$_2$ (Ca$_4$Al$_2$O$_6$Fe$_2$As$_2$) was found to have three (two) Fermi surfaces around $\Gamma$.
We found that the systems of the same crystal structure but different chemical compositions give rise to similar band structures.
By constructing the maximally localized Wannier functions 
and analyzing their transfer integrals,
we demonstrated that the differences between the band structures of Ca$_4$Al$_2$O$_6$Fe$_2$P$_2$ and Ca$_4$Al$_2$O$_6$Fe$_2$As$_2$ originate mainly from the difference in pnictogen height.
Specifically, the hybridization between the $p_Z$ orbital and its neighboring $d_{X^2 - Y^2}$ orbitals plays important roles in determining the electronic structures.

We analyzed the electronic structure of Ca$_4$Al$_2$O$_6$Fe$_2$AsP by unfolding its electronic bands.
We found that its band structure resembles the band structures of Ca$_4$Al$_2$O$_6$Fe$_2$P$_2$ and Ca$_4$Al$_2$O$_6$Fe$_2$As$_2$ with an average crystal structure,
although Ca$_4$Al$_2$O$_6$Fe$_2$AsP contains pnictogen height disorder.
We therefore expected that an accurate calculation of band structures for an arbitrary $x$ is possible using a crystal structure modeled by linearly connecting the crystal structures of Ca$_4$Al$_2$O$_6$Fe$_2$P$_2$ and Ca$_4$Al$_2$O$_6$Fe$_2$As$_2$.

We then calculated the generalized susceptibility $\chi_0$ for Ca$_4$Al$_2$O$_6$Fe$_2$(As$_{1-x}$P$_x$)$_2$ using crystal structures constructed from the linear interpolation of the crystal structures of Ca$_4$Al$_2$O$_6$Fe$_2$P$_2$ and Ca$_4$Al$_2$O$_6$Fe$_2$As$_2$.
It was found that a smaller crystal structure ratio $\lambda$ leads to a larger $\chi_0$ because of the larger density of states and the more effective $\gamma$ Fermi surface.

Further theoretical and experimental investigations of the structural and magnetic properties of perovskite-type systems are required to clarify and optimize the superconducting properties of such systems.

\begin{acknowledgement}
We thank Professor K. Terakura for discussions.
We also thank
Dr. A. Iyo, Dr. H. Eisaki, Dr. C. H. Lee, Dr. P. M. Shirage, Dr. K. Kihou, and Dr. H. Kito
for discussions and for providing us with experimental data prior to publication.
This work was partly supported by the Strategic International Collaborative Research Program (SICORP), Japan Science and Technology Agency,
by the Next Generation Super Computing Project, Nanoscience Program,
and by a Grant-in-Aid for Scientific Research on Innovative Areas,
"Materials Design through Computics: Complex Correlation and Non-Equilibrium Dynamics" (No. 22104010) from MEXT, Japan.
The calculations were performed using computational facilities at TACC, AIST as well as 
at the Supercomputer Center of ISSP and at the Information Technology Center, both at the University of Tokyo. 
\end{acknowledgement}

\end{document}